\documentclass[12pt,preprint]{aastex}

\shorttitle{AGN Infrared number counts}
\shortauthors{Treister et al.}

\begin{document}

\title{Spitzer Number Counts of AGN in the GOODS fields}
\author{Ezequiel Treister\altaffilmark{1,2,3}, C. Megan Urry\altaffilmark{2,4}, Jeffrey Van Duyne\altaffilmark{2},
Mark Dickinson\altaffilmark{5}, Ranga-Ram Chary\altaffilmark{6}, David M. Alexander\altaffilmark{7}, Franz Bauer\altaffilmark{7}, Priya Natarajan\altaffilmark{2}, Paulina Lira\altaffilmark{3}, Norman A. Grogin\altaffilmark{8}}

\altaffiltext{1}{Department of Astronomy, Yale University, P.O. Box 208101,New Haven,CT 06520.}
\altaffiltext{2}{Yale Center for Astronomy \& Astrophysics, Yale University,
P.O. Box 208121, New Haven, CT 06520}
\altaffiltext{3}{Departamento de Astronom\'{\i}a, Universidad de Chile, Casilla 36-D, Santiago, Chile.}
\altaffiltext{4}{Department of Physics, Yale University, P.O. Box 208120, New Haven, CT 06520.}
\altaffiltext{5}{National Optical Astronomy Observatory, Tucson, AZ 85719}
\altaffiltext{6}{Spitzer Science Center, California Institute of Technology, MS 220-6, Pasadena, CA 91125.}
\altaffiltext{7}{Institute of Astronomy, Madingley Road, Cambridge CB3 0HA, UK.}
\altaffiltext{8}{Department of Physics and Astronomy, Johns Hopkins University, Charles and 34th St., Baltimore, MD 21218.}

\email{treister@astro.yale.edu}

\begin{abstract}
We present mid-infrared observations of active galactic nuclei (AGN)
in the Great Observatories Origins Deep Survey (GOODS) fields,
performed with the Spitzer Space Telescope.  These are the deepest
infrared and X-ray fields to date and cover a total area of $\sim$0.1
square degrees. AGN are selected on the basis of their hard (2-8 keV)
X-ray emission. The median AGN infrared luminosity is at least 10
times larger than the median for normal galaxies with the same
redshift distribution, suggesting that the infrared emission is
dominated by the central nucleus. The X-ray to infrared luminosity
ratios of GOODS AGN, most of which are at 0.5$\la$$z$$\la$1.5, are
similar to the values obtained for AGN in the local Universe. The
observed infrared flux distribution has an integral slope of $\sim$1.5
and there are 1000 sources per square degree brighter than
$\sim$50~$\mu$Jy at $\sim$3-6~microns. The counts approximately match
the predictions of models based on AGN unification, in which the
majority of AGN are obscured. This agreement confirms that the
faintest X-ray sources, which are dominated by the host galaxy light
in the optical, are obscured AGN. Using these Spitzer data, the AGN
contribution to the extragalactic infrared background light is
calculated by correlating the X-ray and infrared catalogues. This is
likely to be a lower limit given that the most obscured AGN are missed
in X-rays. We estimate the contribution of AGN missed in X-rays, using
a population synthesis model, to be $\sim$45\% of the observed AGN
contribution, making the AGN contribution to the infrared background
at most $\sim$2-10\% in the 3-24 micron range, depending on
wavelength, lower than most previous estimates. The AGN contribution
to the infrared background remains roughly constant with source flux
in the IRAC bands but decreases with decreasing flux in the MIPS
24~$\mu$m band, where the galaxy population becomes more important.
\end{abstract}


\section{Introduction}

Hard X-ray surveys are not strongly affected by dust
obscuration and thus provide a relatively unbiased view of
the AGN phenomenon. Deep Chandra X-ray surveys have allowed
us to test the basic prediction of population synthesis
models of the X-ray background, namely, that a combination of
obscured and unobscured AGN is needed to explain the X-ray
background hard spectral shape
\citep{setti89,gilli01}. According to the AGN unification
paradigm, obscuration comes from optically thick dust
blocking the central engine along some lines of sight. The
temperature in this structure, which can range up to 1000~K
(dust sublimation temperature), and the roughly isotropic
emission toward longer wavelengths should make both obscured
and unobscured AGN very bright in the mid-far infrared
bands.

This prediction was tested by IRAS (e.g.,
\citealp{rush93}) and ISO (e.g.,
\citealp{fadda02,matute02,franceschini02}) observations, which
discovered a population of very luminous infrared sources 
coincident with X-ray emitters and therefore very good candidates to
be obscured AGN. With relatively bright flux limits, however, these
surveys sample the nearby Universe or extremely luminous, rare
AGN. Very deep surveys were possible with ISO (e.g.,
\citealp{fadda02}), but only over a very small area. With the Spitzer
Space Telescope, it is now possible to perform very deep infrared
surveys, $\sim$100-1000 times deeper than typical IRAS observations,
over appreciable areas, $\sim$10-1,000 times more area than the deep
ISO surveys, thus permitting us, for the first time, to study a
representative sample of the AGN population at
cosmologically-significant distances.

Since AGN comprise most of the X-ray background and are also very
bright infrared sources, it is natural to ask what fraction of the
extragalactic infrared background they may contribute. Based on
previous observations, the predicted contribution of AGN to the
infrared background ranged from $\sim$20\% \citep{fabian99} to just a
few percent \citep{xu01}, depending mostly on the assumptions for the
amount of energy re-radiated in the infrared and the spatial density
of obscured AGN. The deeper infrared observations possible with
Spitzer are important for understanding the true contribution of AGN
to the extragalactic infrared background light.

Multiwavelength observations of a large, uniformly selected
sample of moderate-luminosity AGN at cosmologically
significant distances provide a complete picture of the
AGN emission across the frequency spectrum and establish the
total contribution of these sources to the energy budget of
the Universe. This is one of the main goals of the Great
Observatories Origins Deep Survey (GOODS), a
0.1-square-degree study divided in two fields that overlap
with the deepest X-ray observations, the Chandra Deep Fields
North \citep{brandt01} and South \citep{giacconi01}. It
consists of deep imaging in the optical with the Hubble
Space Telescope \citep{giavalisco04} and in the infrared
with the Spitzer Space Telescope \citep{dickinson02}. These
very deep multiwavelength observations allow a highly
complete, relatively unbiased, view of the AGN phenomenon
and a test of the AGN unification paradigm, well established
in the local Universe, at higher redshifts.

In this paper we present Spitzer observations of the AGN in
the GOODS fields, from 3.6 to 24 microns. In \S~2 we
describe the Spitzer IRAC and MIPS observations. Results and
discussion are presented in \S~3, addressing the infrared
luminosities of AGN and normal galaxies, the X-ray to
infrared ratios for GOODS and local AGN, the infrared number
counts of AGN, and the AGN contribution to the extragalactic
infrared background light. Our conclusions are presented in
\S~4. Throughout this paper we assume $H_0=70$ km s $^{-1}$
Mpc$^{-1}$, $\Omega_m=0.3$ and $\Omega_\Lambda =0.7$.

\section{Spitzer Data and AGN Sample}

The GOODS North and South fields, which cover $\sim$160 square
arcminutes each, were observed with the Spitzer Space Telescope as
part of the GOODS Legacy program (PI: Mark Dickinson). Observations
were performed using the IRAC camera \citep{fazio04a} in the four
available bands (3.6-8$\mu$m) and the MIPS instrument \citep{rieke04}
in the 24~$\mu$m band. Exposure times per sky-pixel for each field
were $\sim$23 hours in the IRAC bands. The MIPS 24-micron
observations, available to us only in the GOODS-North field, had an
exposure time of $\sim$10.3 hours. Individual images were processed
through the Spitzer Science Center basic calibrated data pipeline, and
then post-processed and combined using customized routines developed
by the GOODS team\footnote{More details about data reduction can be
found at the webpage
http://data.spitzer.caltech.edu/popular/goods/Documents/goods\_dataproducts.html}
(M. Dickinson et al., in prep). Sources were detected using SExtractor
\citep{bertin96} on a combined 3.6 and 4.5 $\mu$m band image filtered
using a 5$\times$5 pixel Gaussian profile with a FWHM of
1.8$''$. Photometry was performed using SExtractor. The AUTO magnitude
was used since aperture corrections, which we do not apply, would be
very small, according to simulations\footnote{See the webpage
http://www.noao.edu/noao/staff/ecm/simulations.html for more details
and calculation of aperture corrections and completeness simulations.}
(M. Dickinson et al., in prep.). If instead we perform aperture
photometry (with an aperture diameter of 4$''$) and make the
corresponding aperture corrections, which range from 0.3 magnitudes in
the 3.6 $\mu$m band to 0.7 at 8 $\mu$m then the AUTO and
aperture-corrected magnitudes differ by less than 0.2 magnitudes on
average, with no obvious dependence on magnitude and no significant
outliers. Therefore our results do not depend significantly on the
method used to perform the photometry.

The formal $5\sigma$, background-limited flux limits for an isolated
point source in the four IRAC channels are 0.13, 0.22, 1.45, and 1.61
$\mu$Jy respectively. In practice, the photometric uncertainty for a
given source depends on the degree of crowding in the images, and on
the details of how the photometry and source deblending are carried
out.  This increases the photometric errors, which we evaluate using
Monte Carlo simulations of artificial sources added to the real
data. According to simulations, all the IRAC images are 90\% complete
at $\sim$14~$\mu$Jy.

The MIPS 24-$\mu$m image of the GOODS North field was reduced using a
combination of the standard Spitzer Science Center pipeline and
customized routines created by the GOODS team (R. Chary et al., in
prep.). Photometry was obtained at the IRAC-derived positions using a
point spread function fitting technique.  The 5$\sigma$ flux limit for
the MIPS observations is $\sim$25~$\mu$Jy and are 84\% complete at
that flux.  The relatively large aperture used to calculate fluxes in
both the IRAC and MIPS observations means that the light from the AGN
and the host galaxy are summed; in any case, given the limited spatial
resolution, the two contributions cannot be separated from the
infrared imaging alone.

In order to define the AGN sample we start with the X-ray catalogs for
the Chandra Deep Fields North and South of \citet{alexander03}, with
the associations to the GOODS HST optical catalog from \citet{bauer04}
which uses the likelihood method described by
\citet{bauer00}. Specifically, our sample consists of sources detected
in the hard (2--8 keV) Chandra band that do not have multiple optical
counterparts in the GOODS ACS $z$-band image. Only 3 sources in the
South field and 6 in the North have multiple optical counterparts, so
excluding these sources from the sample does not bias our results. (We
do include sources with no $z$-band counterpart down to the limit
$z$$\simeq$27~mag.) A large fraction of the sample, $\sim$60\%, have
spectroscopic redshifts, obtained by \citet{szokoly04} in the South
field and by \citet{barger03} in the North. Most of the X-ray sources
are obscured AGN at moderate redshift, 0.5$<z<$1, and unobscured AGN
up to $z\sim$4. More distant obscured AGN are too faint for optical
spectroscopy \citep{treister04}. Sources with spectroscopic redshifts
and hard X-ray luminosity lower than $10^{42}$~ergs~s$^{-1}$ are
probably not AGN dominated, since the most X-ray luminous star-forming
galaxy known, NGC 3256, has a total X-ray luminosity $L_X\simeq
8\times 10^{41}$~ergs~s$^{-1}$ in the 0.5-10 keV band \citep{lira02};
so these were removed from our sample. This criterion removes 12
sources in the South field and 25 in the North. With these
constraints, the final sample includes 126 X-ray sources in the
GOODS-South field and 179 in the -North field.

\begin{figure}
\figurenum{1}
\plotone{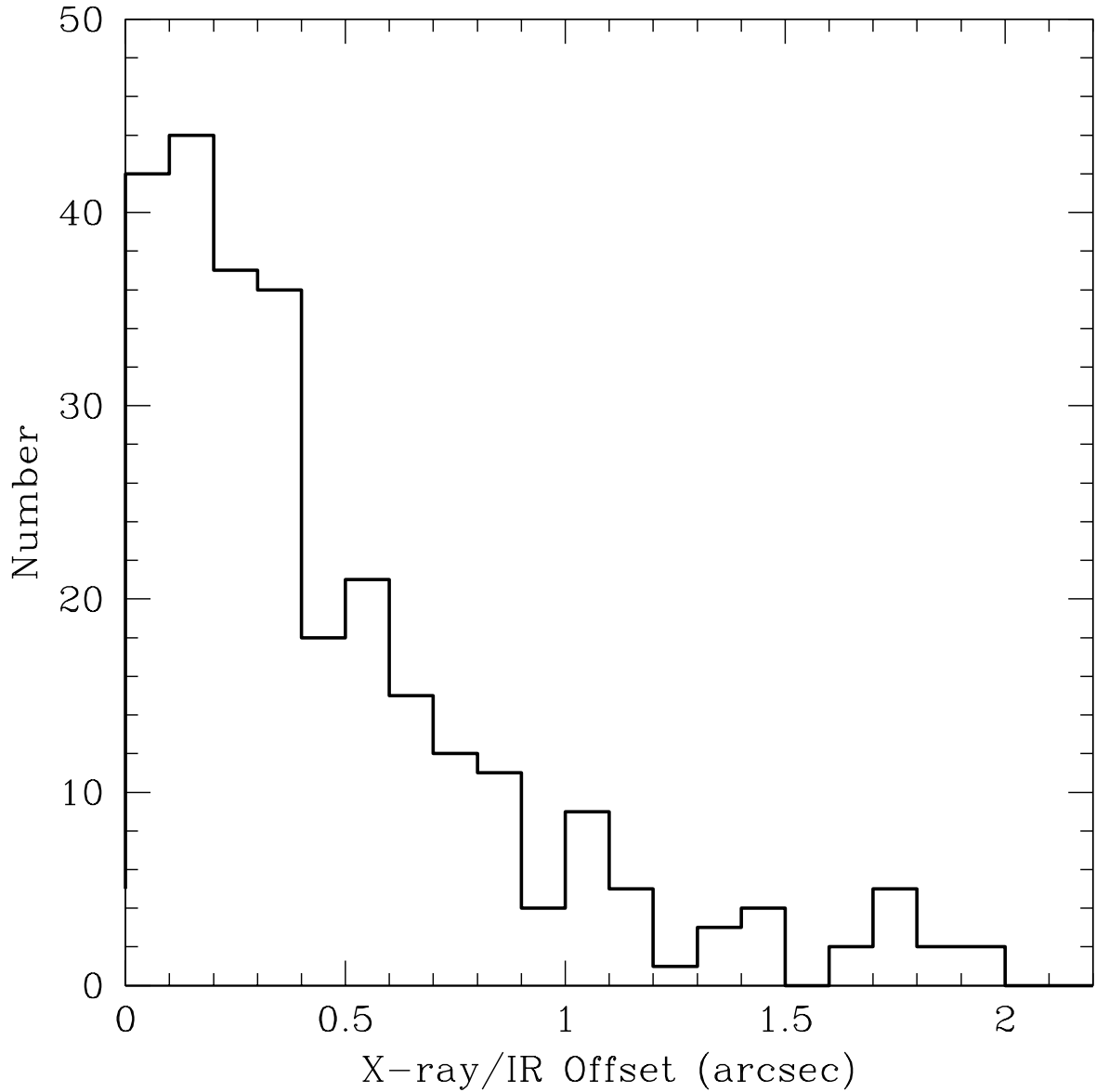}
\caption{Distribution of offsets between the X-ray and infrared 
positions for sources in our AGN sample. Most of the X-ray sources
have a infrared counterpart closer than 1$''$, while for a large
number of AGN, the distance between the X-ray and infrared positions
is smaller than 0.5$''$. Therefore, we are very confident that the
right infrared counterpart for the X-ray sources in our sample was
used in all but one or two cases.}
\label{off_dist}
\end{figure}

In order to look for Spitzer counterparts of the Chandra sources,
several search radii were tried, looking to maximize the number of
sources with counterparts while keeping a low number of sources with
multiple counterparts. The best results were obtained using a search
radius of 2$''$. Typical astrometric errors are $\sim$0.5$''$ in
X-rays and $\sim$0.2$''$ in the IRAC bands; therefore, astrometric
errors can be safely neglected when using a search radius of 2$''$. In
the North field 160 sources (90\%) have a single IRAC counterpart,
while in the south field 118 sources (94\%) have a single counterpart;
the rest have multiple counterparts. {\it This means that 100\% of the
X-ray sources are also bright in the near-infrared}. Most of the X-ray
sources have an infrared counterpart separated by less than 1$''$, and
in many cases this distance is smaller than 0.5$''$, as shown in
Fig.~\ref{off_dist}. Assuming a Poissonian spatial distribution of
infrared sources, the median chance of a random association is 0.74\%
and therefore we have high confidence that the correct infrared
counterpart was identified correctly in all but one or two cases. The
full catalog of AGN detected in the GOODS field used in this work is
presented in the on-line version of the Journal, while for clarity a
fraction of the information is presented in Table~\ref{catalog}.

Our sample only considers AGN selected based on the X-ray
information. Because of this, highly obscured AGN in which most of the
X-ray emission is absorbed will be missed in our sample. The effects
of this incompleteness are discussed in \S 4. Recently, AGN selection
methods based only on IRAC and MIPS colors (e.g., \citealp{sajina05})
or a combination of Spitzer infrared and radio properties (e.g.,
\citealp{martinez05}) were presented. In the work of \citet{sajina05},
AGN are found in the region defined by
0.0$\la$$\log(S_{4.5}/S_{3.6})$$\la$0.3,
0.2$\la$$\log(S_{5.8}/S_{3.6})$$\la$0.5,
0.3$\la$$\log(S_{8}/S_{3.6})$$\la$0.7,
0.1$\la$$\log(S_{5.8}/S_{4.5})$$\la$0.3,
0.3$\la$$\log(S_{8}/S_{4.5})$$\la$0.5 and
0.1$\la$$\log(S_{8}/S_{5.8})$$\la$0.3 using only the IRAC bands
(Fig. 8 on \citealp{sajina05}). Although this criterion was defined
only for AGN at $z$$<$1, K corrections are very small so this method
can be extended to higher redshifts. Only 79 sources satisfy the AGN
criterion of \citet{sajina05} in both the GOODS North and South
fields. Of these, only 18 ($\sim$6\%) sources are detected in hard
X-rays. This means that using this selection method misses a large
fraction of the AGN in the field, but it is useful to separate AGN
continuum-dominated sources from normal galaxies. This can be
explained because a power-law continuum is assumed for the infrared
emission in AGN, while it is possible that most sources are dominated
by the torus thermal emission in the infrared (Van Duyne et al., in
prep).

The selection method of \citet{martinez05} is based mostly in radio
emission and it is effective to detect luminous obscured
quasars. Using the infrared criterion of \citet{martinez05} on the
GOODS North field, where MIPS data were available, only 3 sources have
$S_{24}$$>$300~$\mu$Jy and $S_{3.6}$$<$45~$\mu$Jy and are detected in
hard X-rays. Unfortunately, neither spectroscopic nor photometric
redshifts are available for those sources, so we cannot (yet) classify
them as obscured quasars. The fact that only a small number of sources
are selected using this criterion is not surprising since only very
luminous sources are found in this way and such sources are very rare
and thus not easily detected in narrow-field surveys like GOODS.

\section{Results and Discussion}

\subsection{Infrared Luminosity and Colors}

When spectroscopic redshifts were not available, we used photometric
redshifts computed by \citet{mobasher04} for the South field and
\citet{barger03} for the North field. Using both spectroscopic and
photometric redshifts, we were able to calculate luminosities for 93
sources (74\%) in the South field and 105 sources (59\%) in the North
field. In order to study the quality of photometric redshifts we
compared them with spectroscopic redshifts for sources in which both
measurements are available. The average value of the quantity
($z_\textnormal{spec}$-$z_\textnormal{phot}$)/(1+$z_\textnormal
{spec}$) is 0.02, which indicates that photometric redshifts can be
used in a statistical sense to study the properties of a sample of
sources. However, errors in the photometric redshifts for individual
sources can be large. About 30\% of the sources compared here had a
difference between photometric and spectroscopic redshift
determination greater than 0.3. Therefore, photometric redshifts are
used only to calculate statistical properties of a sample, not for
individual sources.

\begin{figure}
\figurenum{2}
\plottwo{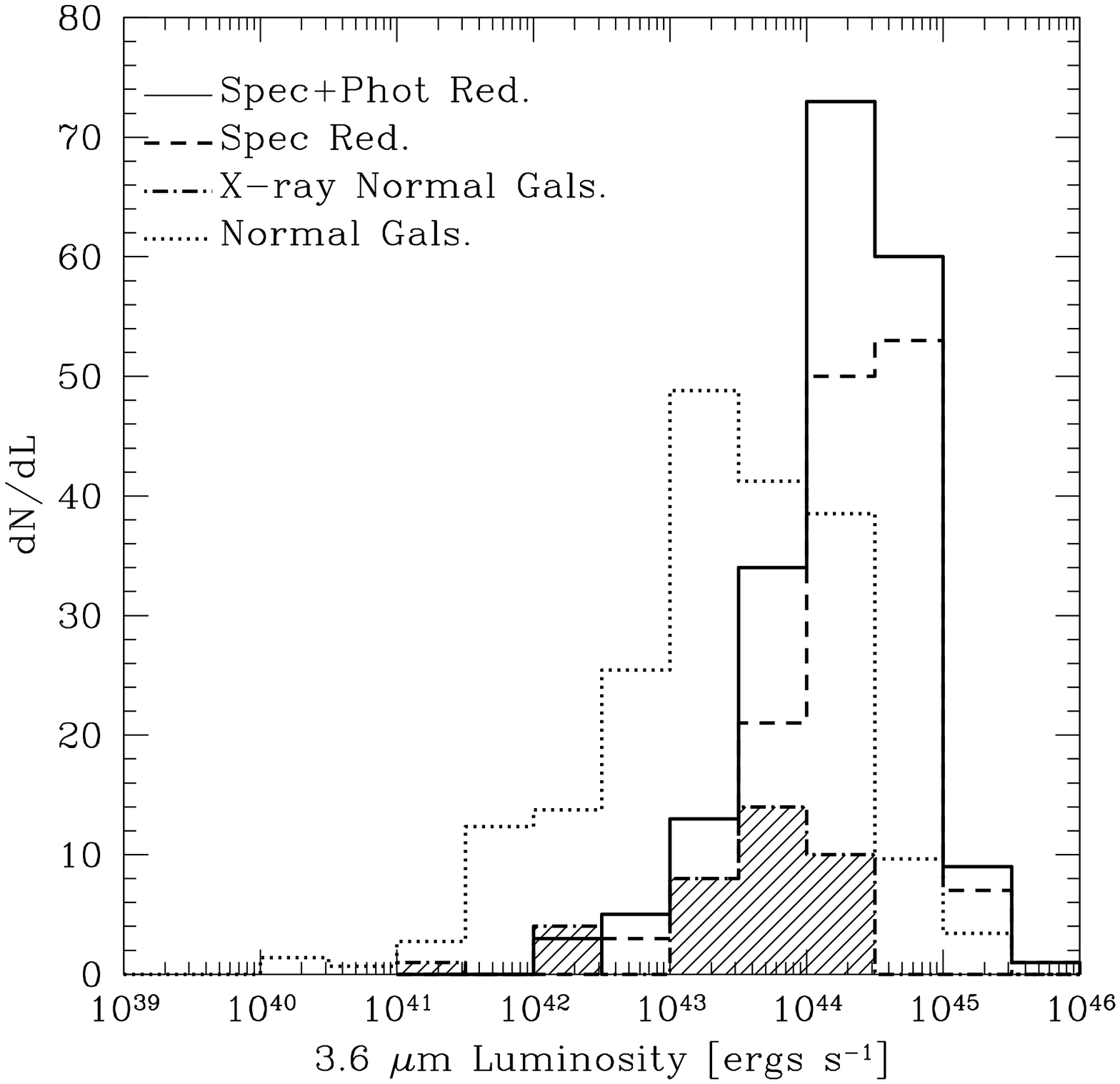}{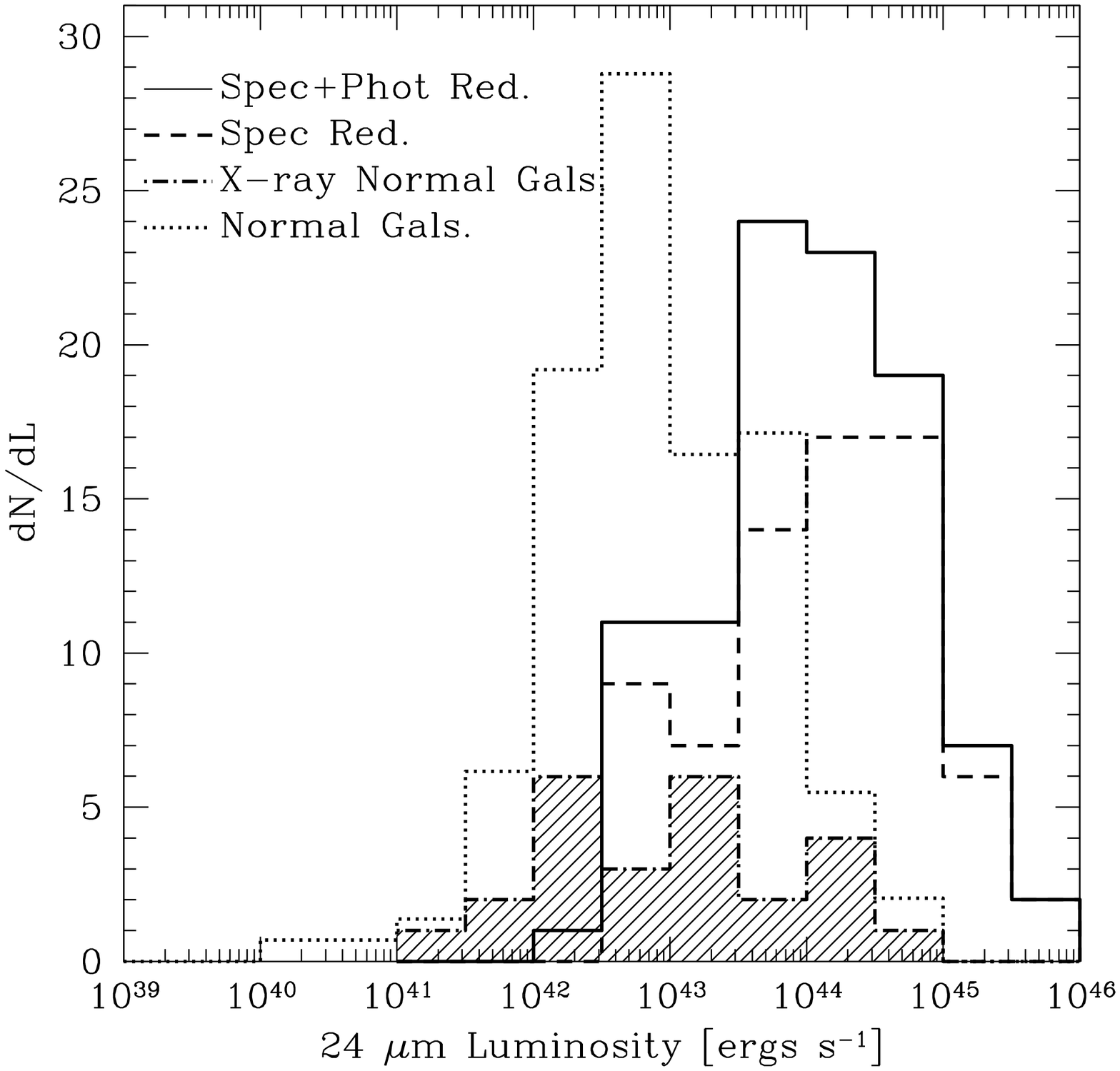}
\caption{Infrared luminosity distribution for AGN
with spectroscopic ({\it dashed lines}) and both spectroscopic or
photometric redshifts ({\it solid lines}) in the IRAC 3.6~$\mu$m ({\it
left panel}) and MIPS 24~$\mu$m ({\it right panel}) bands. Most of the
sources in our sample can be classified as moderate-luminosity AGN, in
the 10$^{42}$-10$^{44}$~ergs~s$^{-1}$ range. {\it Dotted lines:}
Luminosity distribution for random GOODS field galaxies with either
photometric or spectroscopic redshifts selected to have the same
redshift distribution as the AGN sample. Sources classified as AGN in
X-rays are $\sim$10-15 times brighter in the infrared than normal
galaxies. {\it Dot-dashed/hatched histograms:} Luminosity distribution
for X-ray sources with $L_X<$10$^{42}$~ergs~s$^{-1}$, and thus not
included in our AGN sample. These sources have distributions similar
to normal galaxies and much fainter on average than the AGN
luminosity distribution. Therefore, we conclude that the X-ray
emission in these sources does not come from AGN.}
\label{lum_hist}
\end{figure}

The observed-frame luminosity distributions for GOODS AGN in the
IRAC 3.6~$\mu$m and MIPS 24~$\mu$m bands are shown in
Fig.~\ref{lum_hist}. The vast majority of the sources can be
classified as moderate-luminosity AGN, in the range 10$^{42}$ to
$\sim$10$^{44}$~ergs~s$^{-1}$. No K correction was attempted. However,
from the theoretical AGN SED library of \citet{treister04}, which
spans unobscured to fully obscured AGN over 4 decades in luminosity,
we estimate that infrared luminosities in the 3-20 micron range would
change by at most a factor of two due to K corrections. This is
smaller than the bin size used in these histograms, so we do not
expect the figure to change significantly were we able to do accurate
K corrections.

For comparison, in Fig.~\ref{lum_hist} we also show the observed-frame
3.6 and 24 micron luminosity distributions for normal galaxies in the
GOODS fields, randomly selected to follow the same redshift
distribution as the AGN sample. Redshifts for the normal galaxies were
obtained from the spectroscopic surveys of \citet{wirth04} and
\citet{cowie04} in the North field and from the VIMOS VLT Deep Survey
\citep{lefevre04} in the South field. Comparing the median infrared
luminosities of AGN and normal galaxies, AGN are brighter by a factor
of $\sim$10 in the IRAC bands and $\sim$15 in the MIPS 24 $\mu$m
band. There is a fair amount of overlap between the two populations,
indicating that for low-luminosity AGN
($\sim$10$^{42}$--10$^{43}$~ergs~s$^{-1}$) the host galaxy can provide
a significant contribution and perhaps even dominate the infrared
emission. Additionally, the luminosity distribution for sources with
hard X-ray luminosity lower than 10$^{42}$~ergs~s$^{-1}$ (and thus
removed from our AGN sample) is also shown in
Fig.~\ref{lum_hist}. These sources have a luminosity distribution
similar to that of normal galaxies and therefore we conclude that the
X-ray emission in those sources is not from AGN origin and most likely
from star-formation activity.

\begin{figure}
\figurenum{3}
\begin{center}
\plotone{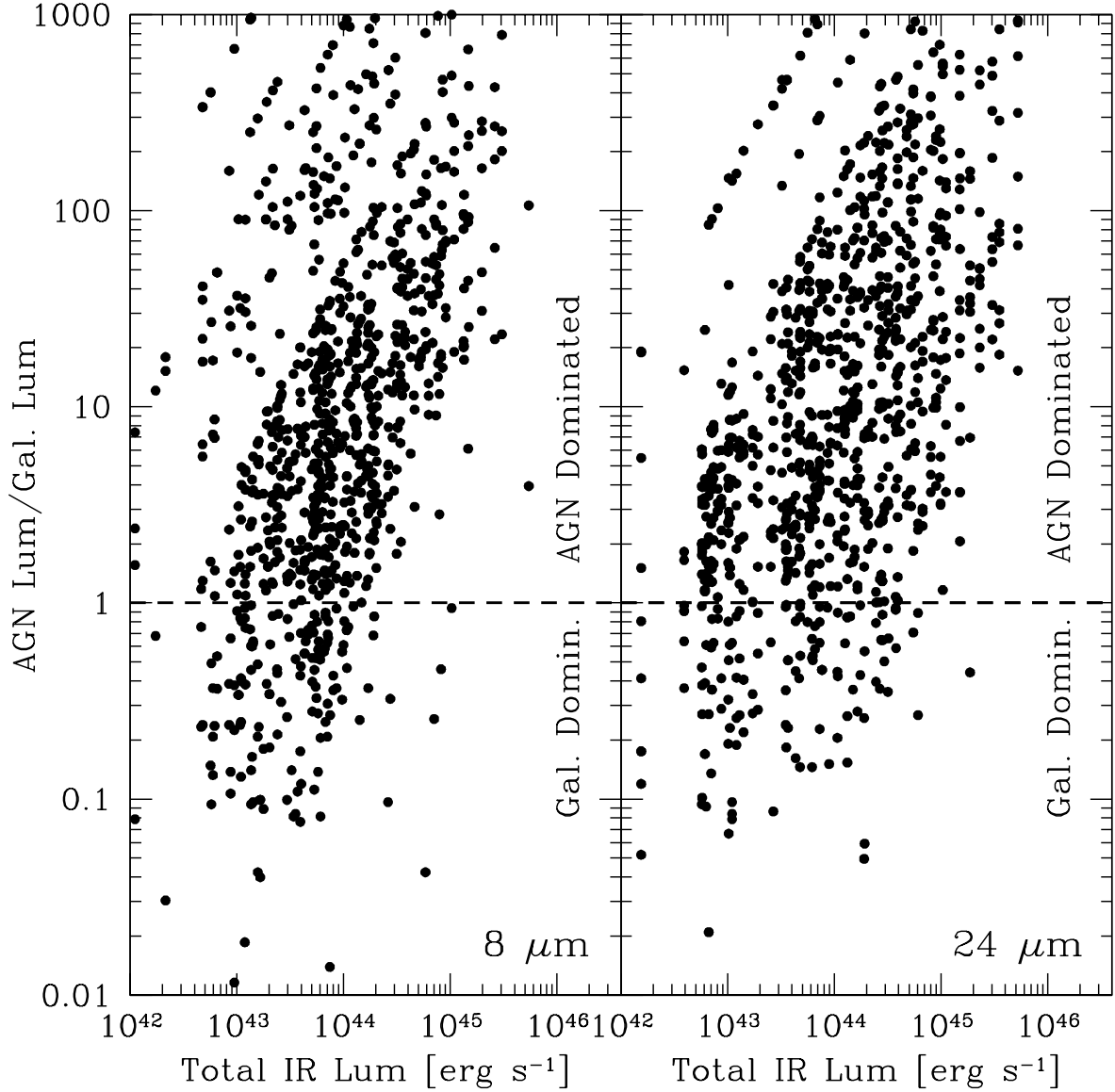}
\end{center}
\caption{Simulated fraction of AGN to host galaxy infrared luminosity as 
a function of total infrared luminosity in the 8 $\mu$m ({\it left
panel}) and 24 $\mu$m ({\it right panel}) bands. The AGN and host
galaxy emission are separated by associating the luminosity of random
galaxies in the field to AGN in our sample. Here the results for 5
random associations for each source are shown. As can be seen, most of
the sources are AGN dominated in the infrared with the fraction
increasing with increasing total luminosity. This result does not
depend much on the observed wavelength.  }
\label{frac}
\end{figure}

In order to understand the relative contributions of infrared emission
from active nucleus and galaxy, we performed a simple simulation in
which we randomly paired a normal galaxy from the redshift-equivalent
distribution (infrared emission from stars only) to each AGN in the
sample (infrared emission from stars plus active nucleus). We then
calculated the distribution of the ratio of galaxy to nuclear light,
assuming the assigned galaxy correctly represented the emission from
the host galaxy.  This process was repeated 150 times with different
random pairings. If indeed these normal galaxies correctly represent
the host galaxy emission, which is of course uncertain but which is
certainly true at optical wavelengths (e.g.,
\citealp{urry99,dunlop03}), the ensuing distribution indicates the
relative importance of the two components. The results of a set of
these simulations are shown in Fig.~\ref{frac} for the 8 and 24 micron
bands. We find that at 8~$\mu$m, $\sim87$\% of the sample is dominated
by the AGN (i.e., the luminosity of the active nucleus is greater than
the galaxy luminosity).  Separating the sample in three luminosity
bins, each with equal numbers of AGN, we find that for total AGN
luminosities below $4\times10^{43}$~ergs~s$^{-1}$, roughly 3/4 of the
sample is AGN-dominated. For $4\times10^{43} < L_{IR} < 1.6 \times
10^{44}$~ergs~s$^{-1}$, the fraction of AGN-dominated sources is
roughly 90\%, and for higher luminosities it rises to 97\%.  At
shorter wavelengths and at 24~$\mu$m, the fractions are similar. At
24~$\mu$m, $\sim$80\% of the sources are AGN-dominated in the lowest
luminosity bin and $\sim$90\% at the highest luminosities. Of course,
because of the limited spatial resolution, we cannot say from the
infrared data alone whether the AGN host galaxies are similar to
normal galaxies in the same redshift range.  Given the similarity of
the optical properties, however, it would require a fairly contrived
scenario to have the infrared properties be massively mismatched. The
correlation of infrared and X-ray flux (next Section) also supports
the idea that the infrared emission from AGN is dominated by the
active nucleus.

\subsection{X-ray to Infrared Ratios}

\begin{figure}
\figurenum{4}
\begin{center}
\plottwo{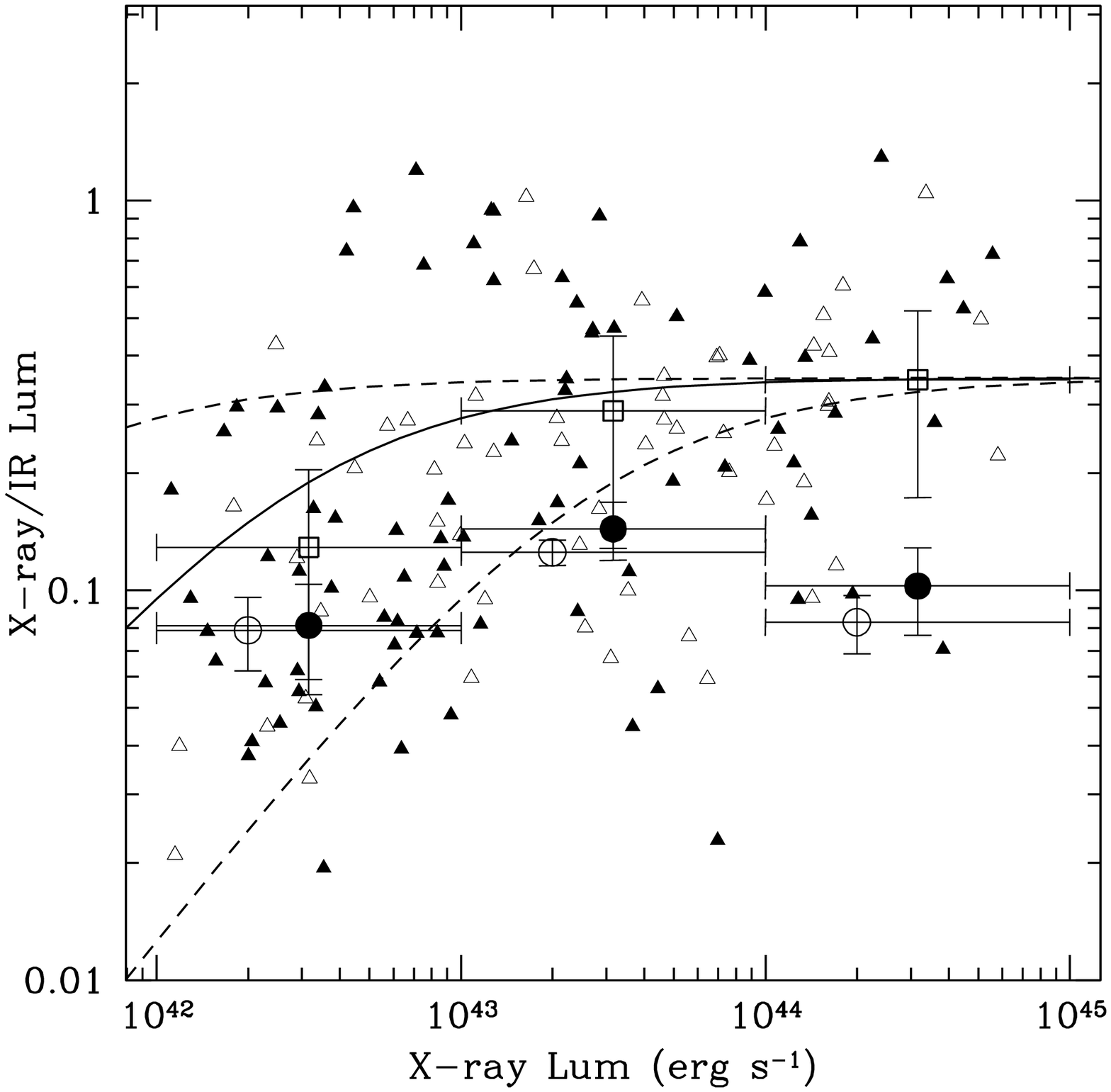}{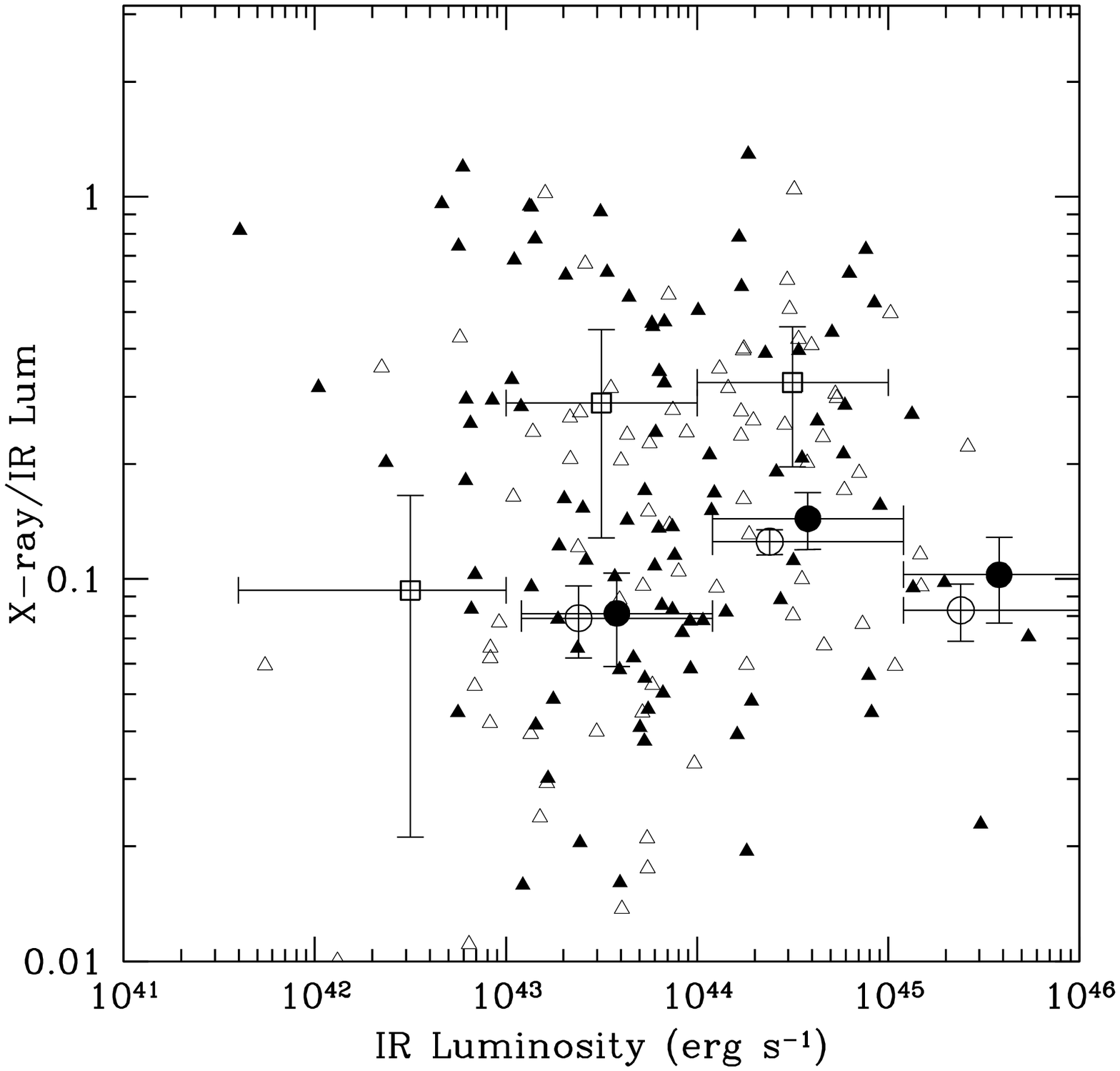}
\end{center}
\caption{{\it Left panel}: Hard X-ray (2-8 keV) 
to infrared (8$\mu$m) luminosity ratio as a function of hard X-ray
luminosity for obscured ({\it solid triangles}) and unobscured ({\it
open triangles}) AGN (using $N_H$=10$^{22}$~cm$^{-2}$ as the
separating point) with measured spectroscopic redshifts in the GOODS
fields. Average values (excluding the top and bottom 10\%) in three
X-ray luminosity bins are shown by the open squares, with error bars
corresponding to the RMS dispersion. For comparison, the average
values for this ratio obtained from a compilation of local AGN by
\citet{silva04} are also shown ({\it filled circles}:
obscured AGN, {\it open circles}: unobscured AGN). GOODS AGN, the
majority of them at $z\sim$0.5-1 (higher-redshift AGN are too faint
for spectroscopy), and local sources have similar values for this
ratio, possibly diverging at higher luminosity. {\it Solid line}:
X-ray to infrared luminosity ratio assuming an intrinsic ratio of 0.35
and a host galaxy with a luminosity of
7.67$\times$10$^{42}$~ergs~s$^{-1}$ (median luminosity for our random
sample of GOODS normal galaxies). The {\it dashed lines} show the
effects of changing the luminosity of the host galaxy by one order of
magnitude (lower line: brighter host galaxy). The small observed trend
with luminosity can be explained by the relatively more important
contribution from the host galaxy for low-luminosity AGN. {\it Right
panel}: Hard X-ray to infrared luminosity ratio as a function of
infrared luminosity. Symbols are the same as in the left panel. In
this case, sources below the X-ray luminosity threshold are also
included in the diagram. The same conclusions are still valid when the
X-ray to infrared luminosity ratio is measured as a function of
infrared or hard X-ray luminosity.}
\label{x_ir}
\end{figure}

In Fig.~\ref{x_ir} we show the ratio of hard X-ray (2--8 keV) to
infrared (8~$\mu$m) luminosity versus hard X-ray and infrared
luminosity for GOODS AGN with spectroscopic redshifts and thus
reliable luminosities. Using the X-ray spectrum to separate obscured
and unobscured AGN (the separation was set at
$N_H$=10$^{22}$~cm$^{-2}$, with $N_H$ inferred from the X-ray hardness
ratio, assuming an intrinsic photon slope of $\Gamma$=1.7), we find no
differences between the two classes of AGN (filled and open triangles
in Fig.~\ref{x_ir}). This is consistent with the AGN unification
paradigm since both hard X-ray and infrared emission are
\textit{relatively} isotropic and unaffected by the amount of dust in
the line of sight. Similar results are found if the 24 micron fluxes
are used instead; however, these fluxes are only available for half of
our sample and thus we decided to use the 8 micron fluxes.

In order to compare with previously known sources, in Fig.~\ref{x_ir}
we also show the average observed-frame hard X-ray to infrared
luminosity ratio for both local Seyfert 1 and 2 galaxies in three
X-ray luminosity bins, from the compilation of \citet{silva04}. The
average ratio for the local sources is 0.10$\pm$0.02 (standard
deviation for the sample, including K correction), considering both
obscured and unobscured AGN and excluding the top and bottom 10\%
values of the distribution. For the GOODS sample this ratio is
0.18$\pm$0.12, also excluding the top and bottom 10\% values, and
there is also no significant difference in the average values for the
obscured and unobscured AGN. The average value for the GOODS sample in
three X-ray luminosity bins is shown by the open squares in
Fig.~\ref{x_ir}. There is no statistically significant discrepancy
between the GOODS and local samples, although there is a hint that the
values might diverge at high X-ray luminosity ($<$2$\sigma$
significance). However, it should be noted that Malmquist bias,
which would make the average luminosity to appear to be higher at
higher redshifts, can play a role in this difference.

The trend for the average ratio of X-ray to infrared
luminosity to increase with X-ray luminosity in the GOODS
sample can be explained by the contribution of the host
galaxy to the infrared light at low luminosities. Assuming
that this ratio has a value of 0.35 for the highest
luminosity sources, which are clearly AGN-dominated, and
adding to the infrared luminosity the contribution from an
average host galaxy (estimated from the median 8~$\mu$m
luminosity for our random sample of galaxies,
7.67$\times$10$^{42}$~ergs~s$^{-1}$), we were able to
reproduce the observed trend of X-ray to infrared
luminosity ratio with X-ray luminosity (solid line in
Fig.~\ref{x_ir}).

The spread in the values of the X-ray to infrared luminosity for the
GOODS sample can be explained by the combined effects of ignoring the
K correction and the intrinsic spread in the luminosity of the host
galaxy. Using the AGN spectra library of \citet{treister04}, we
estimate that the K correction can account for variations of $\sim
50$\% around the mean observed ratio. The dashed lines in
Fig.~\ref{x_ir} show the effects of a spread in host galaxy
luminosity, which is particularly important for low-luminosity AGN and
can account for most of the observed spread in the X-ray to infrared
luminosity ratio at these luminosities.

Performing a partial correlation analysis we found no significant
correlation between the hard X-ray to infrared luminosity ratio and
either hard X-ray or infrared luminosity, or with redshift or
absorbing column density ($N_H$). We found only a mildly significant
correlation between hard X-ray and infrared luminosities, with a
Pearson coefficient of $\sim$0.604 and a correlation of similar
significance between hard X-ray and infrared fluxes, indicating that
the fluxes in these bands are directly related to the luminosity of
the central engine. {\it The lack of correlation of the ratio with
redshift or luminosity suggests there is little dependence of the
emission mechanism on either redshift or luminosity.}

\subsection{Infrared Counts of AGN}

\begin{figure}
\figurenum{5}
\plotone{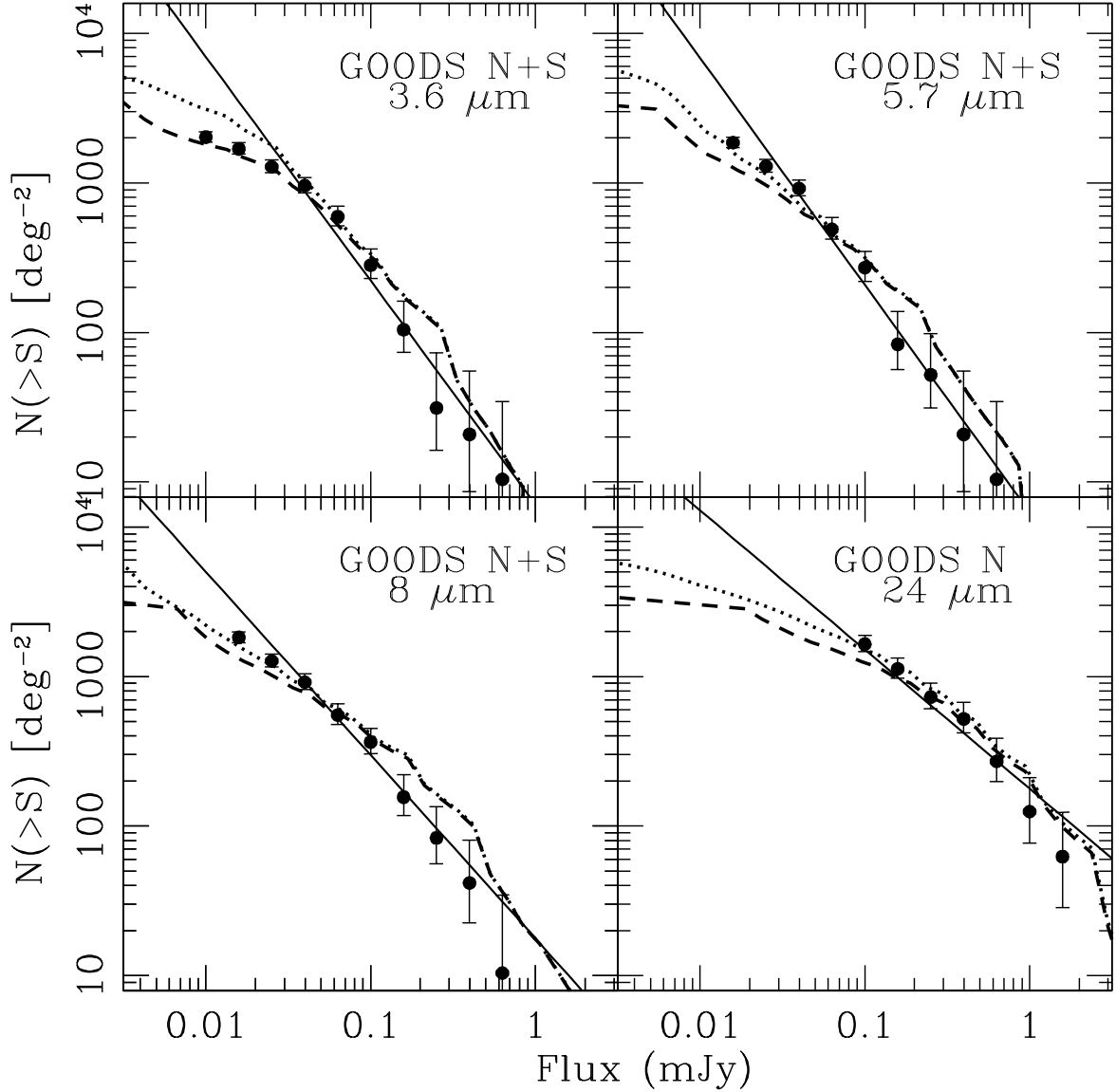}
\caption{{\it Filled circles:} Observed cumulative infrared 
flux distributions for X-ray detected AGN in the GOODS North and South
fields in three IRAC bands (3.6-8$\mu$m) and for the North field only
in the MIPS 24$\mu$m band. Error bars correspond to the 84\%
confidence level in the number of sources on that bin
\citep{gehrels86}. Note that because this is a cumulative plot, the
bins are not independent. {\it Solid line:} Power-law fit to the
observed number counts. {\it Dashed line:} Expected infrared flux
distribution based on the models of \citet{treister04}, taking into
account the X-ray flux limit in the GOODS fields and correcting for
the effects of undersampling of the bright end caused by the small
volume of the GOODS fields. {\it Dotted line:} Expected infrared flux
distribution not considering the X-ray flux limit (i.e., all sources
expected from extrapolating the AGN luminosity function). In general,
the model follows the normalization and shape of the counts reasonable
well, especially considering the poor statistics at the bright end. At
the faint end, where the host galaxy light dominates, the model
diverges more significantly since the model assumptions are clearly
far too simple.}
\label{dist}
\end{figure}

The cumulative distributions of infrared fluxes in the 3.6, 5.7, 8 and
24 micron bands for the GOODS AGN are presented in
Fig.~\ref{dist}. The observed counts were corrected to account for the
differences in covered area depending on the X-ray flux (since the
sample is X-ray selected) based on the relation given by Fig.~5 in
\citet{treister04}, obtained by adapting the sensitivity versus area
function for the Chandra Deep Fields of \citet{alexander03} to the
portion covered by GOODS (D. Alexander and F. Bauer, priv. comm.). No
correction for incompleteness in the X-ray or infrared catalogues is
necessary as these are almost complete in the flux ranges sampled in
the number counts determination. Error bars in these distributions
were calculated following the prescription of \citet{gehrels86} for
84\% confidence levels on the number of sources. The number counts in
the 3.6, 4.5 and 5.7~$\mu$m bands can be well described by a power-law
with integral slope 1.5$\pm$0.2 for fluxes higher than 50~$\mu$Jy from
a maximum likelihood analysis of the unbinned data
\citep{crawford70}. In the 8~$\mu$m band the integral slope is
slightly shallower, 1.2$\pm$0.2, at fluxes higher than 70~$\mu$Jy,
with a density of $\simeq$1000 AGN per square degree at a depth of
$\sim$35~$\mu$Jy. In the MIPS 24~$\mu$m band the slope is shallower
still, 0.9$\pm$0.2, at all the fluxes sampled by the GOODS sources,
and there are 1000 AGN per square degree at a flux density of
$\sim$200~$\mu$Jy. The resulting best fits are shown in
Fig.~\ref{dist}.  In order to test the goodness of these fits we
performed a K-S test comparing the observed distribution to the
fitting function, finding that the null hypothesis (that both
distributions are drawn from the same parent distribution) cannot be
rejected at confidence levels higher than $\sim$40\% in all the
bands. The values of the K-S probability for each band are 45\%, 50\%,
75\%, 85\% and 40\% on the 3.6, 4.5, 5.7, 8 and 24~$\mu$m bands
respectively.

In the unification paradigm for AGN, most of the energy absorbed by
the circumnuclear optically thick matter is re-emitted at infrared
wavelengths. It is not surprising then that AGN are luminous infrared
sources, even when the optical, ultraviolet and soft X-ray emission is
faint. Population synthesis models for the X-ray background suggest
that a substantial fraction of AGN are obscured along the line of
sight, thus will be bright only at mid- to far- infrared and hard
X-ray wavelengths. For a simple unification model, in which the
material is distributed in a torus and the ratio of obscured to
unobscured AGN is set to $\sim$3:1, as observed in the local Universe
\citep{risaliti99}, we can calculate the expected infrared
counts. Model details are as described in \citet{treister04}
and modified by \citet{treister05b} to include a decrease in
the number of obscured AGN with increasing luminosity, with
the exception that we no longer interpolate between the
near-infrared and optical spectra because this artificially
over-predicts the mid-infrared flux. 

The obscured-to-unobscured AGN ratio for sources with spectroscopic
redshifts (and thus $N_H$ determinations) in our sample is
$\sim$1.5:1, but this is consistent with the assumed ratio in the
model, given that most of the obscured AGN do not have spectroscopic
redshift since they are optically faint, and some obscured AGN are
even missed in X-rays. Therefore, the observed value of the ratio only
represents a lower limit to the intrinsic ratio, which can easily be
higher by a factor of two and therefore consistent with the average
value assumed in our model.

This simple AGN unification model agrees within a factor of two with
the observed distribution in the GOODS fields over the IRAC 3.6-8
micron and MIPS 24 micron bands, as shown in Fig.~\ref{dist}. Compared
to the model, which assumes only an elliptical host galaxy, with a
single ($L_*$) luminosity and no star formation, and fixed parameters
for the dust torus, there are slightly fewer observed AGN at the
bright end ($>$100~$\mu$Jy). The small area of the GOODS survey causes
bright AGN to be poorly sampled, hence the large error bars and the
marginal significance of the discrepancy. Additionally, it is
important to point out that in a cumulative plot the bins are not
independent. Wide area surveys like SWIRE \citep{lonsdale03} and the
MIPS GTO\footnote{More information can be found at the web site
http://lully.as.arizona.edu/} programs will provide a better estimate
of the AGN counts at the bright end. At the faint end ($<$10~$\mu$Jy),
the observed number of AGN is higher than the model predictions except
in the 3.6~$\mu$m band where there is very good agreement. These
discrepancies are $\sim$13\% at 4~$\mu$m and rising to 36\% at
8~$\mu$m and $\sim$30\% at 24~$\mu$m. These discrepancies may suggest
that for fainter AGN, the host galaxy plays a relatively more
important role in some AGN, thus making them brighter. At 3.6~$\mu$m
the stellar component in the host galaxy dominates; the good agreement
with our simple model means that the host galaxy must be reasonably
well represented by our assumed $L_*$ elliptical galaxy with no recent
star formation.

\subsection{Extragalactic Infrared Background Light}

\begin{figure}
\figurenum{6}
\includegraphics[scale=0.7]{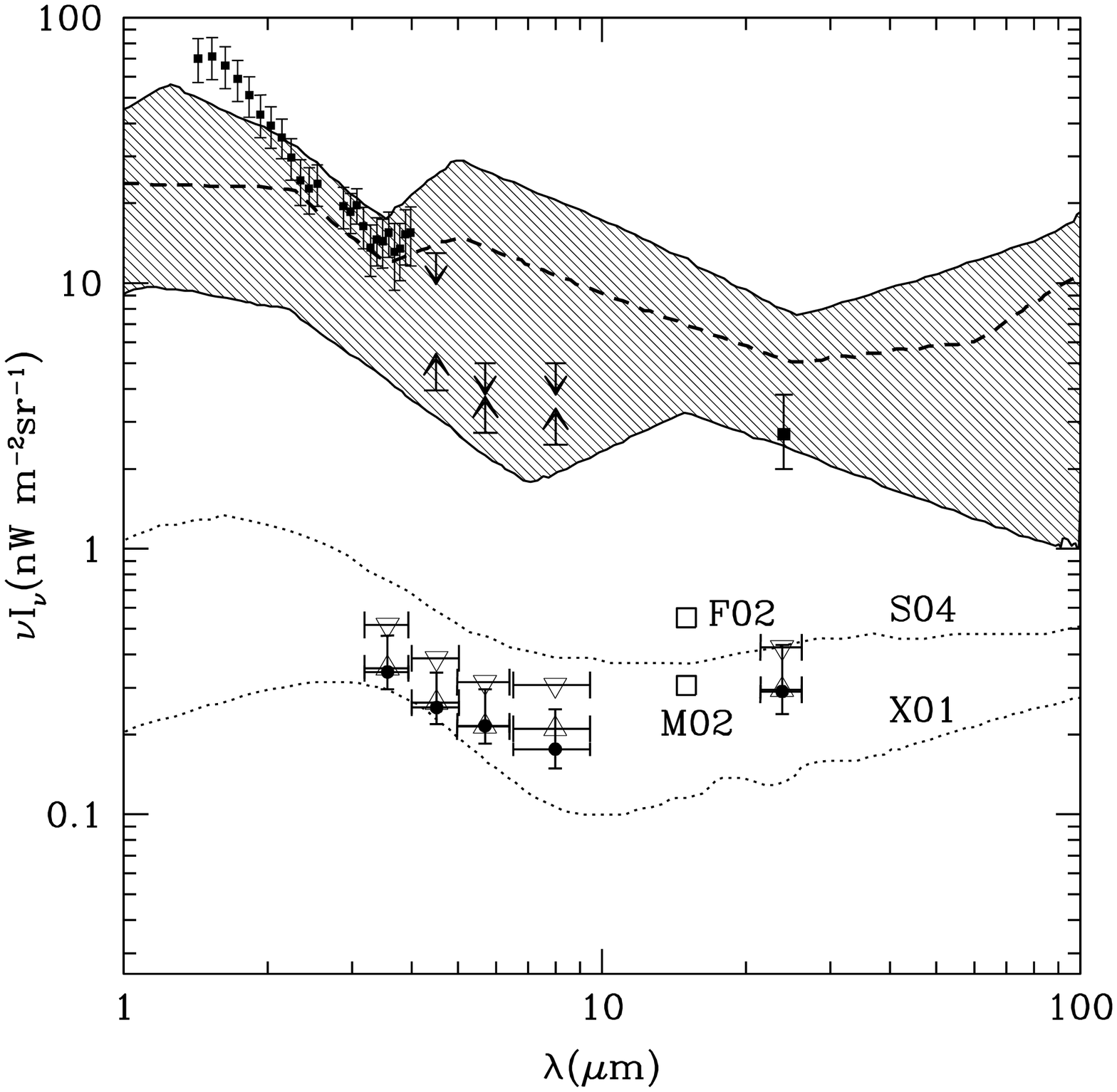}
\caption{Extragalactic infrared background intensity as a function of 
wavelength. {\it Shaded region}: Allowed values compiled by
\citet{hauser01}. 1-4~$\mu$m: Measurements reported by
\citet{matsumoto05}. 4-8$\mu$m: Lower limits from
\citet{fazio04b}, upper limits from \citet{kashlinsky96} at
4.5~$\mu$m and \citet{stanev98} at 5.8 and
8~$\mu$m. Measurement at 24~$\mu$m from \citet{papovich04}.
{\it Solid circles}: Integrated infrared emission from X-ray
detected AGN in the GOODS fields. Error bars in these
measurements are estimated based on the number of
sources. {\it Triangles}: Integrated AGN emission from the
models of \citet{treister04} and
\citet{treister05b} if only X-ray detected AGN are
considered ({\it lower}) and including all AGN ({\it
upper}). {\it Dotted lines:} Expected AGN integrated
emission from the AGN models of Silva et al. (2004; {\it
upper line}) and Xu et al. (2001; {\it lower line}). {\it
Open squares}: Expected AGN integrated emission at 15$\mu$m
from the extrapolation to fainter fluxes of IRAS and ISO
(\citealp{matute02}; M02; corrected by a factor of 5 to include
the contribution from obscured AGN) and ISO only
(\citealp{fadda02}; F02) observations.}
\label{back} 
\end{figure}

Using the Spitzer data we can estimate the integrated contribution of
AGN to the extragalactic infrared background light at IRAC and MIPS
wavelengths. A representative lower limit can be obtained simply by
summing the infrared emission from the AGN detected in the GOODS
fields, weighted by the effective area at that X-ray flux. In
Fig.~\ref{back}, we show the resulting contribution measured from the
GOODS sources, compared to the observational limits for the total
extragalactic infrared background compiled by Hauser \& Dwek (2001;
the lower limit is from the integrated extragalactic light, while the
upper limit comes from fluctuations analysis) and updated measurements
at the IRAC and MIPS 24~$\mu$m wavelengths
\citep{matsumoto05,fazio04b,kashlinsky96,stanev98,papovich04}. AGN and
their host galaxies contribute approximately 2.4$_{-0.7}^{+1.7}$\%,
3.0$_{-1.3}^{+5.7}$\%, 5.6$_{-1.9}^{+5.3}$\%, 4.7$_{-1.7}^{+5.4}$\% in
the IRAC 3.6, 4.5, 5.7 and 8 $\mu$m bands respectively, and
10.7$_{-4.5}^{+11.0}$\% in the MIPS 24-micron band. (The uncertainties
are dominated by the range in the total infrared background light.) 
These are lower limits to the total AGN contribution, since they
exclude the contribution from faint sources below the GOODS X-ray
detection limit (including a large expected population of
Compton-thick AGN; \citealp{treister05b}) and from bright sources that
are not well sampled in the relatively small GOODS fields. Using our
AGN unification model, we can calculate the relative contribution of
these X-ray faint sources and add it to the total emission from X-ray
detected sources, in order to estimate the total contribution of AGN
to the infrared background. We find that sources missed in X-rays can
add $\sim$45\% more flux in the 3--24 micron range, with $\sim$30\%
extra flux coming from obscured AGN that are too faint to be detected
in X-rays and the remaining $\sim$15\% coming from bright sources that
are not well sampled in the GOODS fields. Correcting for this factor,
the AGN contribution is increased to $\sim$3-8\% in the IRAC bands and
$\sim$15\% in the MIPS 24-micron band. Given that these values are
obtained from the summed emission of AGN {\it and} their host
galaxies, these are upper limits to the AGN contribution to infrared
background, or equivalently, can be seen as the contribution to the
infrared background by AGN plus their host galaxies.

\begin{figure}
\figurenum{7}
\plotone{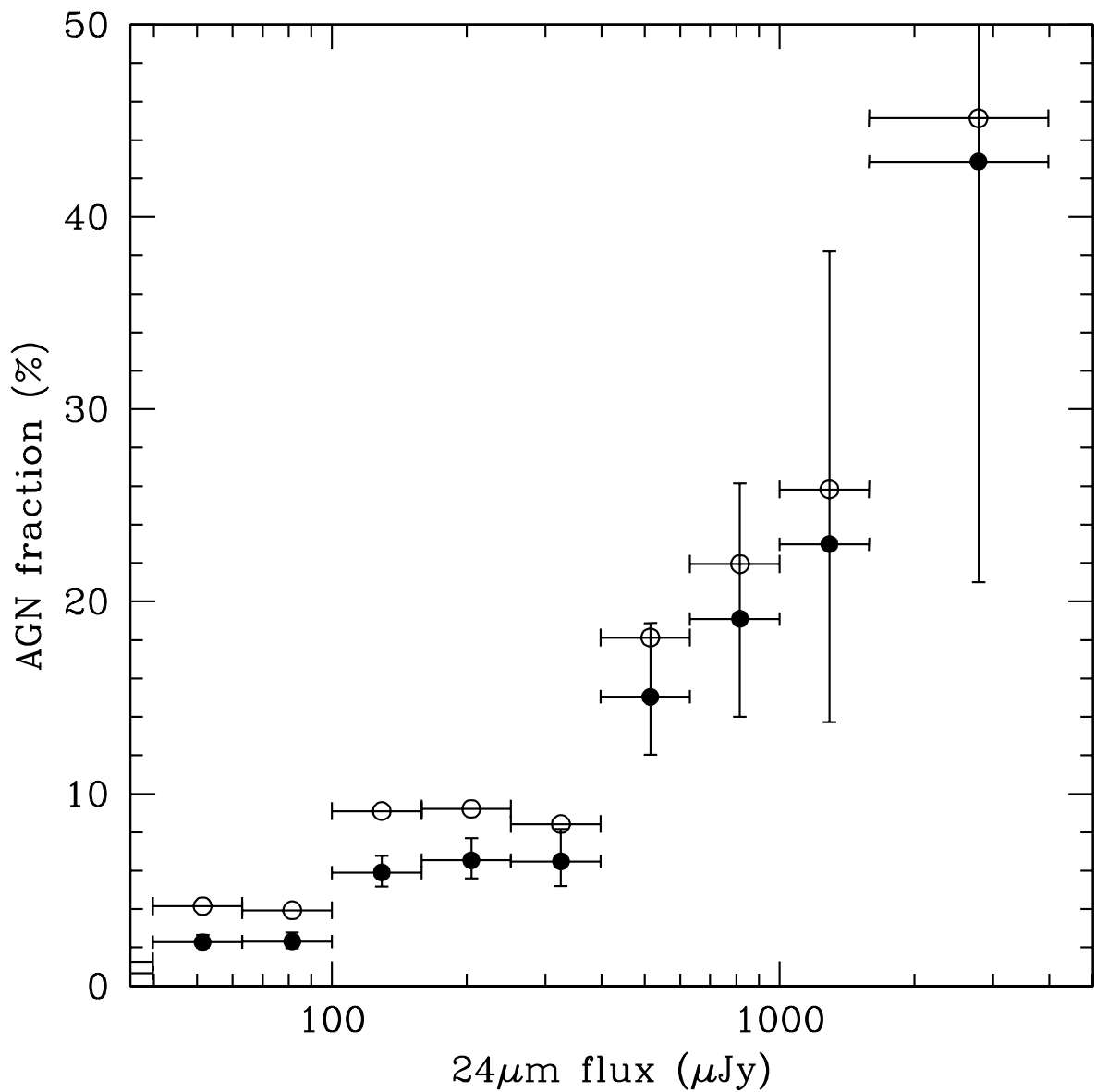}
\caption{Fraction of sources classified as AGN increases with infrared 
flux in the 24~$\mu$m band ({\it solid circles}). Horizontal error
bars show the size of the flux bins, while vertical error bars show
the 1$\sigma$ Poissonian errors on the number of sources. In order to
test if this effect can be caused by the X-ray selection, we also show
the contribution corrected by the AGN expected to be missed by X-ray
selection, as estimated using the AGN population synthesis model ({\it
open circles}). As can be seen, this effect is still observed after
correcting for the AGN not detected in X-rays.}
\label{agn_frac}
\end{figure}

We also investigate the flux dependence of the AGN contribution to the
integrated infrared emission. This was computed by comparing the total
flux from AGN with the integrated flux from all sources for several
flux bins. At 3.6 microns the AGN contribution is 2.1\% above
600~$\mu$Jy, rising to $\sim$4\% above 10~$\mu$Jy. At 8 microns, the
AGN contribution is $\sim$6\% and remains mostly constant, with a
slight increase to $\sim$10\% at 40~$\mu$Jy. Only in the MIPS
24~$\mu$m band do we see a decrease in the AGN contribution with
decreasing flux, going from 30\% above 1000~$\mu$Jy to $\sim$8\% above
100~$\mu$Jy, as shown in Fig.~\ref{agn_frac}. In other words, at lower
fluxes the number of galaxies rises much more steeply than the number
of AGN. While this can also mean that many faint AGN are missed by the
X-ray selection, in Fig.~\ref{agn_frac} we use the previously
described AGN model to estimate the AGN fraction as a function of
infrared flux, correcting for AGN missed by the X-ray selection. As
can be seen in Fig.~\ref{agn_frac}, the decrease in the AGN fraction
with decreasing infrared flux is still observed, even after correcting
for AGN missed in X-rays.

Previous reports of the AGN contribution in general
estimated a higher contribution to the integrated
emission. Earlier infrared observations with ISO (e.g.,
\citealp{fadda02}) in the 15-micron band at brighter fluxes
give higher values for the AGN contribution (17\% at
15$\mu$m) but depend on large extrapolations to lower
fluxes. (The GOODS MIPS observations are $\sim$25 times
deeper than the ISO data on the Lockman Hole.) 
\citet{matute02} made an estimate based on ISO observations
of bright unobscured AGN in the ELAIS S1 field. If, in order to
calculate the total AGN contribution from their measured value for
unobscured AGN we assume an obscured-to-unobscured ratio of $\sim$4:1,
their results are consistent with the observed values obtained from
the GOODS observations. An estimate based on X-ray samples and an
average X-ray-to-infrared ratio by \citet{silva04} is $\sim$50-80\%
higher than the values obtained directly from the Spitzer
observations.  Using a backward evolution model based on IRAS and ISO
observations and infrared luminosity functions, the AGN contribution
to the infrared background was also calculated by \citet{xu01}. In
this case only an infrared color criterion was used to identify
AGN. This prediction is similar to the observed values in the 3.6 and
4.5~$\mu$m bands but estimates a lower contribution at longer
wavelengths, $\sim$55\% lower at 24~$\mu$m. A recent study of AGN
detected by Chandra using Spitzer performed by
\citet{franceschini05}, reported a value for the AGN
contribution to the infrared background of 10\%-15\% at
24~$\mu$m in good agreement with our results.

The earlier estimates of the AGN contribution to the
integrated extragalactic infrared light are therefore
largely consistent with our results; the earlier numbers
differ mainly because of large extrapolations to lower
fluxes. Summarizing, the more precise values from our deep
Chandra and Spitzer observations are 2.4\% at 3.6~$\mu$m,
3.0\% at 4.5~$\mu$m, 5.6\% at 5.7~$\mu$m, 4.7\% at 8~$\mu$m
and 10.7\% at 24~$\mu$m. These numbers should be multiplied
by a factor of $\sim$1.45 in order to account for the
contribution by Compton-thick AGN not detected in X-rays by
Chandra, namely, 3.5\% at 3.6~$\mu$m, 4.4\% at 4.5~$\mu$m,
8.1\% at 5.7~$\mu$m, 6.8\% at 8~$\mu$m and 15.5\% at
24~$\mu$m.

\section{Conclusions}

We present here Spitzer observations of the
moderate-luminosity, moderate- to high-redshift AGN in the
GOODS fields, which include the deepest infrared and X-ray
observations to date. In general these sources have similar
X-ray to infrared luminosity ratios as AGN in the local
Universe. AGN are $\sim$10-15 times brighter in the infrared
than normal galaxies, although there is some overlap between
the luminosity distribution of AGN and normal galaxies,
indicating that for low-luminosity AGN the contribution to
the infrared light from the host galaxy can be very
important and may well dominate the observed emission. This
can explain the observed trend of the X-ray-to-infrared
luminosity ratio to increase with X-ray luminosity.

We report the observed infrared flux distribution for GOODS
AGN, which has a slope of $\simeq$1.5 to 0.9 for fluxes
higher than $\sim$100~$\mu$Jy, with a source density of
$\simeq$1000 AGN per square degree at a depth of $\sim$50
$\mu$Jy in the IRAC bands. Comparing to the expectations of
a unification model for AGN, we find reasonable agreement,
showing that the basic properties of the moderate to high
redshift AGN population are well explained by the same
unification paradigm that fits local sources very well, but
with a factor of two less dust in the circumnuclear region.

From the Spitzer observations, we calculate the minimum AGN
contribution to the extragalactic near-mid infrared background,
obtaining a lower value than previously estimated, ranging from 2\% to
10\% in the 3-24 micron range. Accounting for heavily obscured AGN
that are not detected in X-ray, the AGN contribution to the infrared
background increases by $\sim$45\%, to $\sim$3-15\%.

\acknowledgments

We thank the GOODS team for very useful comments and the anonymous
referee for suggestions that improved the paper.  ET thanks the
support of Fundaci\'on Andes, Centro de Astrof\'{\i}sica FONDAP and
the Sigma-Xi foundation through a Grant in-aid of Research. This work
was supported in part by NASA grant HST-GO-09425.13-A.

\clearpage\thispagestyle{empty}
\begin{deluxetable}{cccccccccccccc}
\tablecolumns{14}
\tabletypesize{\scriptsize}
\rotate
\tablewidth{0pc}
\tablecaption{Catalog of AGN in the GOODS fields.}
\tablehead{
\colhead{Field} & \colhead{ID} & \colhead{Prob.\tablenotemark{a}} & \multicolumn{2}{c}{X-ray Flux (erg~cm$^{-2}$~s$^{-1}$)} &\colhead{Hardness Ratio\tablenotemark{b}} &\multicolumn{5}{c}{Infrared Flux ($\mu$Jy)} & \colhead{Redshift} & \colhead{Redshift type\tablenotemark{c}} & \colhead{Classification\tablenotemark{d}} \\
\colhead{} & \colhead{} & \colhead{} & \colhead{Soft (0.5-2 keV)} & \colhead{Hard (2-8 keV)} & \colhead{} & \colhead{3.6~$\mu$m} & \colhead{4.5~$\mu$m} & \colhead{5.7~$\mu$m} & \colhead{8~$\mu$m} & \colhead{24~$\mu$m} & \colhead{} & \colhead{} & \colhead{}
}
\startdata
S &34 & 0.011 & 3.58e-15 &7.03e-15& -0.42&  24.89&  26.55 & 30.76 &  33.73  & -1.00& 1.037 &S& B\\
S &39 & 0.007 & 2.77e-15 &5.62e-15& -0.41&   8.24&   8.47 &  9.55 &  14.32  & -1.00& 2.726 &S& B\\
S &41 & 0.004 & 4.35e-15 &8.61e-15& -0.42&  52.97&  60.26 & 59.70 &  53.95  & -1.00& 1.616 &S& B\\
S &42 & 0.032 & 2.04e-16 &6.74e-16& -0.24&   5.06&   6.49 &  9.46 &   5.81  & -1.00& 9.999 &N& U\\
S &43 & 0.015 & 6.84e-16 &8.99e-16& -0.54&   7.18&   7.94 &  6.25 &   6.73  & -1.00& 2.573 &S& N\\
\enddata
\tablenotetext{a}{Probability of random association assuming a Poissonian
            distribution of sources. Expression 1-$\exp$(-n $\pi$ $d^2$) was used
            where n is the density of sources and $d$ is the distance between
            the X-ray and infrared sources.}
\tablenotetext{b}{Defined as (H-S)/(H+S) where H and S are the background-substracted counts in the Hard and Soft X-ray bands respectively.}
\tablenotetext{c}{S: Spectroscopic Redshift, P: Photometric Redshift, N: No redshift measured}
\tablenotetext{d}{B: Broad line AGN, N: Narrow line AGN, G: Galaxy, U: Unknown}
\tablecomments{This table is published in its entirety in the electronic edition of the Astrophysical Journal. A portion is shown here for guidance regarding its form and content.}
\label{catalog}
\end{deluxetable}

\clearpage

\end{document}